\documentclass[english,longauth]{aa}
\usepackage{epsf,amsfonts,amssymb,graphicx,fancyheadings,caption}
\usepackage{natbib}
\usepackage[latin1]{inputenc}
\usepackage{babel}
\bibpunct{(}{)}{;}{a}{}{,}

\begin{document}

\title{Discovery of a peculiar Cepheid-like star towards the northern edge of the Small Magellanic Cloud
\thanks{Based on observations made by the EROS-2 collaboration with the MARLY, 1.54 m Danish and 3.60 m telescopes at the European
Southern Observatory, La Silla, Chile.}}

\author{
J.B.~Marquette\inst{1},
P.~Tisserand\inst{2,3},
P.~François\inst{4}\fnmsep\thanks{\emph{Present address: }European Southern Observatory (ESO), Casilla 19001, Santiago 19, Chile},
J.P.~Beaulieu\inst{1},
V.~Doublier\inst{5},
É.~Lesquoy\inst{2,1},
A.~Milsztajn \inst{2}\fnmsep\thanks{deceased},
J.~Pritchard\inst{5},
A.~Schwarzenberg-Czerny\inst{6,7},
C.~Afonso\inst{2}\fnmsep\thanks{\emph{Present address: }Max-Planck-Institut für Astronomie, Koenigstuhl 17, D-69117 Heidelberg, Germany},
J.N.~Albert\inst{8},
J.~Andersen\inst{9},
R.~Ansari\inst{8},
É.~Aubourg\inst{2},
P.~Bareyre\inst{2},
X.~Charlot\inst{2},
C.~Coutures\inst{2,1},
R.~Ferlet\inst{1},
P.~Fouqué\inst{10},
J.F.~Glicenstein\inst{2},
B.~Goldman\inst{2}\fnmsep\thanks{\emph{Present address: }Max-Planck-Institut für Astronomie, Koenigstuhl 17, D-69117 Heidelberg, Germany},
A.~Gould\inst{11},
D.~Graff\inst{11}\fnmsep\thanks{\emph{Present address: }Division of Medical Imaging Physics, Johns Hopkins University, Baltimore, MD 21287-0859, USA},
M.~Gros\inst{2},
J.~Haissinski\inst{8},
C.~Hamadache\inst{2},
J.~de Kat\inst{2},
L.~Le Guillou\inst{2}\fnmsep\thanks{\emph{Present address: }LPNHE, CNRS-IN2P3 and Universités Paris 6 \& Paris 7, 4 place Jussieu, 75252 Paris Cedex 05, France},
C.~Loup\inst{1}\fnmsep\thanks{\emph{Present address: }Observatoire Astronomique de Strasbourg, UMR 7550, 11 rue de l'Universit\'{e}, 67000 Strasbourg, France},
C.~Magneville\inst{2},
\'E.~Maurice\inst{12},
A.~Maury\inst{5}\fnmsep\thanks{\emph{Present address: }San Pedro de Atacama Celestial Exploration, Casilla 21, San Pedro de Atacama, Chile},
M.~Moniez\inst{8},
N.~Palanque-Delabrouille\inst{2},
O.~Perdereau\inst{8},
Y.R.~Rahal\inst{8},
J.~Rich\inst{2},
M.~Spiro\inst{2},
A.~Vidal-Madjar\inst{1},
S.~Zylberajch\inst{2}
}

\institute{
Institut d'Astrophysique de Paris, UMR 7095 CNRS, Universit\'{e} Pierre \& Marie Curie, 98~bis Boulevard Arago, 75014 Paris, France
\and
CEA, DSM, DAPNIA, Centre d'Études de Saclay, 91191 Gif-sur-Yvette Cedex, France
\and
Research School of Astronomy \& Astrophysics, Mount Stromlo Observatory, Cotter Road, Weston ACT 2611, Australia
\and
GEPI, Observatoire de Paris, 61 avenue de l'Observatoire, 75014 Paris, France
\and
European Southern Observatory (ESO), Casilla 19001, Santiago 19, Chile
\and
Centrum Astronomiczne im. M. Kopernika, Bartycka 18, 00-716 Warszawa, Poland
\and
Obserwatorium Astronomiczne, Uniwersytet A. Mickiewicza, Sloneczna 36, 60-286 Poznan, Poland
\and
Laboratoire de l'Accélérateur Linéaire, IN2P3 CNRS, Universit\'e de Paris-Sud, 91405 Orsay Cedex, France
\and
The Niels Bohr Institute, Copenhagen University, Juliane Maries Vej 30, DK2100 Copenhagen, Denmark
\and
Observatoire Midi-Pyrénées, Laboratoire d'Astrophysique (UMR 5572), 14 av. E. Belin, 31400 Toulouse, France
\and
Department of Astronomy, Ohio State University, Columbus, OH 43210, U.S.A.
\and
Observatoire de Marseille, 2 place Le Verrier, 13248 Marseille Cedex 04, France
}
\offprints{J.B. Marquette; \email{marquett@iap.fr}}

\date{Received ; Accepted}

\abstract
{For seven years, the EROS-2 project obtained a mass of photometric data on variable stars.
We present a peculiar Cepheid-like star, in the direction of  the Small Magellanic Cloud,
which demonstrates unusual photometric behaviour over a short time interval.}
{We report on data of the photometry acquired by the MARLY telescope and spectroscopy
from the EFOSC instrument for this star, called \object{EROS2 J005135-714459(sm0060n13842)},
which resembles the unusual Cepheid \object{HR 7308}.}
{The light curve of our target is analysed using the Analysis of Variance method to determine a
pulsational period of 5.5675 days. A fit of time-dependent Fourier coefficients is
performed and a search for proper motion is conducted.}
{The light curve exhibits a previously unobserved and spectacular change in both mean magnitude
and amplitude, which has no clear theoretical explanation. Our analysis of the spectrum implies a radial velocity
of  104 km s$^{-1}$ and a  metallicity of -0.4$\pm$0.2 dex. In the direction of right ascension, we measure a proper motion of
17.4$\pm$6.0 mas yr$^{-1}$ using EROS astrometry, which is compatible with data from the NOMAD catalogue.}
{The  nature of EROS2 J005135-714459(sm0060n13842) remains unclear. For this star, we may have detected a
non-zero proper motion for this star, which would imply that it is a foreground object. Its radial velocity,
pulsational characteristics, and photometric data, however, suggest that it is instead a Cepheid-like object located in the SMC. In such a case, it would present a challenge to conventional Cepheid models.}


\authorrunning{J.B. Marquette et al.}
\titlerunning{A peculiar Cepheid-like object towards the SMC}

\maketitle

\section{Introduction}
Measuring amplitude and period variations in the light curves of pulsating stars remains an ongoing challenge in studying stellar variability, which requires
repeated observations of many objects.
Following the pioneering research of \citet{1929MNRAS..90...65E} about
period irregularities of long-period variable stars, numerous authors examined particular characteristics of different stellar populations, such as the period fluctuations of both Mira stars by \citet{1999PASP..111...94P} and Cepheids by \citet{1999JAVSO..27....5T}.
\citet{2003PASP..115..626P} showed that the non-evolutionary period changes in the $\beta$ Cephei star \object{BW Vul}
are not the result of random cycle-to-cycle fluctuations.
An interesting detection of amplitude change in the stellar class of Cepheids is that for \object{V473 Lyr}.
\citet{2001MNRAS.322...97K} summarized the various hypotheses of different authors who attempt
to explain these changes in amplitude, and proposed that the Blazhko effect could explain the
observations of V473 Lyr.

Cepheids of the Magellanic Clouds were also studied, because of their importance as standard candles. \citet{1985MNRAS.212..395D} reported data about the period change of 115 Cepheids, while the relations between the period and period change of LMC Cepheids were compared by
\citet{1990PASJ...42..341S}. Based on data from Harvard, OGLE and ASAS, the studies by
\citet{2001AcA....51..247P, 2002AcA....52..177P} reported the period changes for both LMC and SMC
Cepheids. \citet{2003AcA....53...63P} compared period changes,
inferred from O-C diagrams for Galactic Cepheids, and model calculations applying particular emphasis to
the first overtone object \object{Polaris}, for which the period change was far more rapid than had been predicted.

Another interesting example was the $s-$Cepheid \object{HR 7308}, which was recognized in the early
eighties \citep{1980A&A....91..115B}. This star exhibited an amplitude variation of a factor of 6 over
about 1200 days that was reported by \citet{1982A&A...109..258B}, who
concluded that the variation was not due to the beating between two oscillation modes because the shape of the
radial velocity curve indicated the pulsation was radial with a unique frequency.
However, a model was proposed, which involved a resonance between a linearly unstable radial mode
and a linearly stable low-degree, low-order p-mode \citep{1995A&A...295..361V}.

The most striking example of a peculiar object was probably the \object{V19}
variable star discovered by \citet{1926ApJ....63..236H} in M33, which was
classified at that time as a Cepheid; its nature remains a mystery, as discussed by \citet{2001ApJ...550L.159M}.
This object exhibited a 54.7 day period, an intensity-weighted mean $B$ magnitude of 19.6, and a $B$ amplitude of 1.1 mag.
\citet{2001ApJ...550L.159M} concluded that its amplitude was less than
0.05 mag and its mean $B$ magnitude had risen to 19.08 mag.
They concluded that V19 could be a high-mass Population I star of a behaviour similar to
a low-mass Population II RV Tauri variable. They noted that it was desirable to avoid
the misclassification of stars such as V19 as classical Cepheids in distant galaxies.

Large microlensing surveys are particularly efficient detectors of such objects. The main goal of the EROS-2 experiment was to search for microlensing events \citep{1986ApJ...304....1P} due to baryonic dark matter in the Halo \citep{2007A&A...469..387T} or to ordinary stars in the Galactic plane \citep{2001A&A...373..126D, 2006A&A...454..185H}.
These rare events are detected by filtering tens of million light curves, some of which may contain stellar nuggets. We report the discovery, during a search for microlensing
events in an outer region of the SMC, of a Cepheid-like star with peculiar characteristics:
both its mean magnitude and amplitude are decreasing spectacularly with time (see Fig. \ref{LightCurves.fig}). After a brief description of
the basics of the EROS-2 setup (Sect. \ref{sec:setup}), we describe the analysis of its peculiar photometric behaviour in Sect. \ref{sec:photom} and in Sect. \ref{sec:spectro}, we describe the spectroscopic data analysis. The photometric blending issue is discussed in Sect. \ref{sec:blending}.
Before presenting our conclusions, we present the proper-motion measurements obtained for the star EROS2 J005135-714459(sm0060n13842) and the post-AGB hypothesis
in Sect. \ref{sec:prop_mot} and \ref{sec:post-AGB}, respectively.

\section{Photometric setup and data acquisition}
\label{sec:setup}
The photometric data were acquired between July 1996 and February 2003 using the MARLY telescope (1m Ritchey-Chrétien, f/5.14),
with a dichroic beam-splitter; this permitted simultaneous imaging to be completed in two non-standard broad passbands, the
so-called EROS filters $B_{\rm E}$ (4200-7200 \AA{}, ``blue'' channel) and $R_{\rm E}$ (6200-9200 \AA{}, ``red'' channel). Each camera consisted of a mosaic of eight 2K $\times$ 2K LORAL CCDs with a pixel size of 0.6\arcsec ~, and a field of view of 0.7\degr
(right ascension) $\times$ 1.4\degr (declination). The $x$ coordinate on a CCD increases with decreasing
declination ($e.g.$ towards South), while the $y$ coordinate increases with increasing right ascension.

\begin{figure}
	\resizebox{\hsize}{!}{\includegraphics{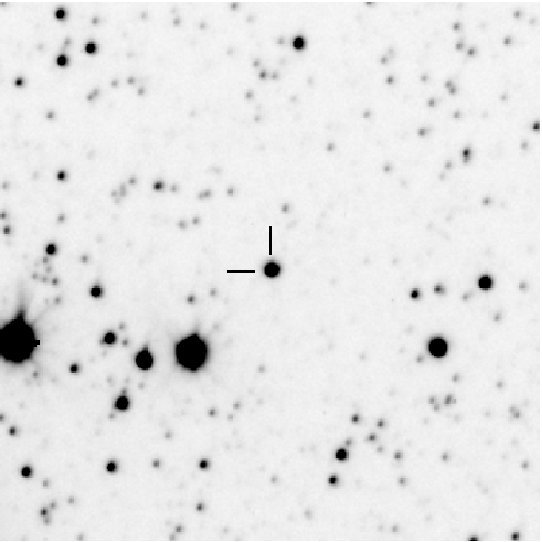}}
	\caption{Finding chart of sm0060n13842, which is indicated by the solid lines. North is up, East is to the left.
The size of the entire square window is 168\arcsec .}
	\label{Champ.fig}
\end{figure}

In the case of the SMC, ten fields were observed. The light curves of individual stars were constructed from fixed positions on templates
using PEIDA, a software dedicated to the photometry of EROS-2 images \citep{1996VA.....40..519A}.
The stars were labelled according to the rules defined for the catalogue by \citet{2002A&A...389..149D}.
In the present paper, we study the star is EROS2 J005135-714459(sm0060n13842), hereafter sm0060n13842, whose J2000
coordinates are 00:51:35.65; -71:44:59.5 (galactic: $l =$ 302.915\degr, $b =$ -45.378\degr). Thus, sm0060n13842 is situated at the northern edge of the SMC. Figure \ref{Champ.fig} provides a finding chart of this object.

\section{Photometric analysis}
\label{sec:photom}
\begin{figure}
	\resizebox{\hsize}{!}{\includegraphics{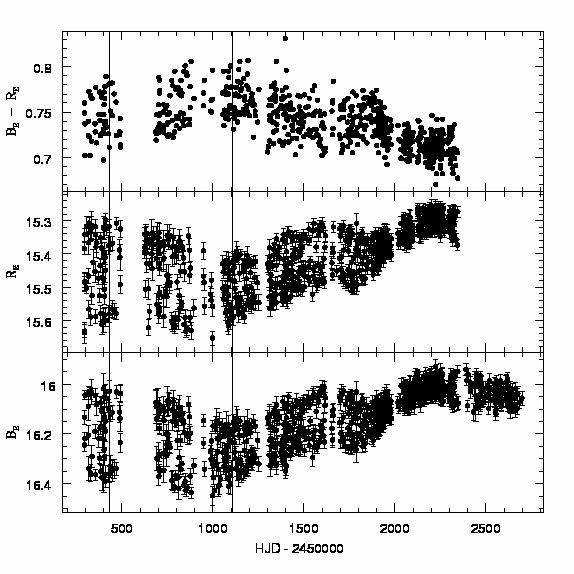}}
	\caption{EROS-2 light curves for sm0060n13842. The left vertical solid line indicates the epoch
of the DENIS observation, while the right one is that of the 2MASS observation. For the sake of clarity, error bars are omitted from the top panel.}
	\label{LightCurves.fig}
\end{figure}

We now examine the photometric properties of sm0060n13842. Figure \ref{LightCurves.fig} shows the EROS-2 light curves\footnote{The photometric data are available in electronic form at the CDS via anonymous ftp to cdsarch.u-strasbg.fr (130.79.128.5) or via http://cdsweb.u-strasbg.fr/cgi-bin/qcat?J/A+A/} for this star in the colours $B_{\rm E}$ and $R_{\rm E}$
with the colour index $B_{\rm E} - R_{\rm E}$. We note that there is no $R_{\rm E}$ data after JD $\sim$ 2\,452\,360 because of technical problems with the camera.
During the EROS-2 observing period it appears that sm0060n13842 exhibited
significant variations in both mean magnitude and amplitude. A slight change is visible in
the colour index. We note that this object is not present in the public MACHO and OGLE-II databases.

Additional $BVRI$ standard photometry was obtained by the PLANET collaboration in August 2004 at
the 1.54 m Danish telescope (ESO, La Silla), which is summarized in Table \ref{Photom.tab}. These data agree well with measurements derived by \citet{2002AJ....123..855Z} using their $UBVI$ catalogue of the SMC. We note that these authors reported sm0060n13842 to be a star whose $UBVI$ photometry cannot be fitted to a stellar atmosphere
model. Additional data from the two infrared surveys DENIS \citep{2000A&AS..144..235C} and 2MASS
\citep{2003yCat.2246....0C} are also indicated.
\begin{table}
	\caption{$BVRI$ photometry of sm0060n13842 obtained at the 1.54 m Danish telescope (ESO, La Silla) at mean JD = 2\,453\,229.74239
(12 August 2004, 05:49 UTC). Values in parentheses are from \citet{2002AJ....123..855Z}.
Additional near-infrared data from DENIS \citep{2005yCat....102002T} and 2MASS \citep{2003yCat.2246....0C}
are also indicated.}
	\label{Photom.tab}
	\medskip
	\centering
	\begin{tabular}{lll}
	\hline
	\hline
	Band		& Magnitude & Julian date\\
	\hline
	$B$		& 17.25 (17.29) &	2\,453\,229.75770 \\
	$V$		& 16.26 (16.21) & 2\,453\,229.74493 \\
	$R$		& 15.61 & 2\,453\,229.73094 \\
	$I$		& 15.11 (15.09) & 2\,453\,229.73600 \\
	$I_{DENIS}$	& 15.07 & 2\,450\,429.57333 \\
	$J_{DENIS}$	& 14.18 & 2\,450\,429.57333 \\
	$J_{2MASS}$	& 14.27 & 2\,451\,106.75840 \\
	$H_{2MASS}$& 13.72 & 2\,451\,106.75840 \\
	$K_{DENIS}$	& 13.66 & 2\,450\,429.57333 \\
	$K_{2MASS}$& 13.50 & 2\,451\,106.75840 \\
	\hline
	\end{tabular}
\end{table}

The {\it local} shape of the light curves is similar to that of a classical Cepheid, while its
global shape certainly not. We understand that there is presently no theoretical interpretation of this behaviour.

\begin{figure}
	\resizebox{\hsize}{!}{\includegraphics{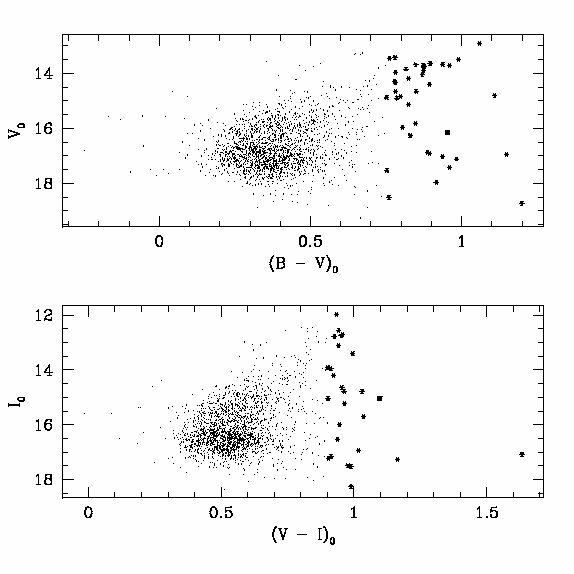}}
	\caption{Dereddened colour-magnitude diagrams $I_0$ versus $(V - I)_0$ and $V_0$ versus $(B - V)_0$
of the SMC Cepheids from OGLE \citep{1999AcA....49..437U}. The filled squares indicate the position of sm0060n13842. For clarity, OGLE points are indicated by stars if $(B - V)_0  > 0.75$ and $(V - I)_0 > 0.90$.}
	\label{CMD.fig}
\end{figure}

Figure \ref{CMD.fig} displays the colour-magnitude diagrams $I_0$ {\it vs} $(V - I)_0$ and $V_0$ {\it vs} $(B - V)_0$ of the OGLE Cepheids in the SMC
\citep{1999AcA....49..437U}.
The filled squares represent the position of sm0060n13842. These values were calculated using the data in
Table \ref{Photom.tab}, and extinction coefficients, in the direction of the SMC, obtainded from the NASA/IPAC Extragalactic Database.

To study the Cepheid nature of the object, we performed a Fourier analysis of its light curve.
It was necessary to remove the long-term variation in magnitude and amplitude since we have insufficient measurements to characterise the variations as, for example, periodic, even though the EROS-2 observation period spanned 7 years. We will consider only the $B_{\rm E}$ data set because it is the most complete. Completeness is important for determining the period because the period accuracy is inversely proportional to both the square root of the number of
measurements (924 in $B_{\rm E}$ and 810 in $R_{\rm E}$) and to the total time of observations
\citep{1985PASP...97..285G}.

We first calculated the mean flux $\overline{F_{\rm lc}}$ and the variance $\sigma^2_{\rm lc}$ for the entire light curve.
We then considered the data to be seasonal (7 seasons were accumulated) and divided each season into 2 parts of
equal numbers of points. For each part, the average date and flux were calculated. We had 2 points per
season from which a linear interpolation was performed to obtain the temporal evolution of the mean flux during a given season,
$\overline{F_{\rm s}(t)}$. We were then able to determine the mean flux for every epoch of the season.

\begin{figure}
	\resizebox{\hsize}{!}{\includegraphics{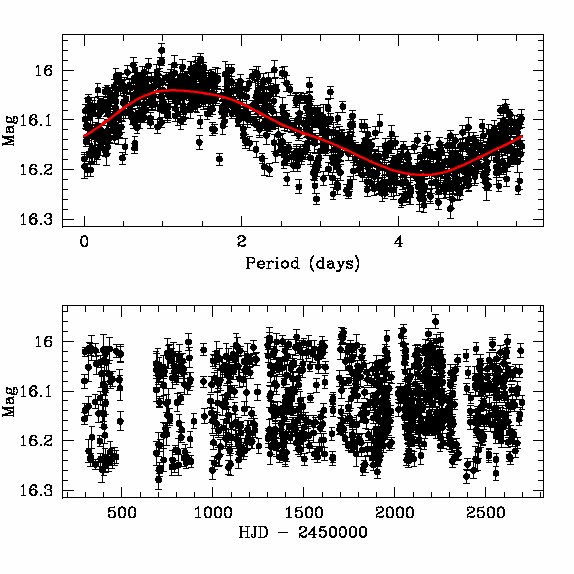}}
	\caption{$B_{\rm E}$ light curve obtained after removing long timescale variations as described in the text. Lower panel:
complete, corrected light curve from $F^*(t)$; upper panel: folded light curve with a period of 5.5675 days. The red line indicates the Fourier fit, the coefficients of which are given in the first row of Table \ref{Fourier.tab}.}
	\label{complete.fig}
\end{figure}

We analysed the variable amplitude by dividing each season into 4 parts. This defined 28 temporal windows, denoted
below by index $j$. For each window we selected the 2 brightest and the 2 faintest measurements. For these 4 points, we calculated the differences $|F_i - \overline{F_{\rm s}(t_i)}|$ for fluxes $F_i$ at times $t_i~(1 \leq i \leq 4)$. These 4 differences were
averaged to obtain $\overline{D_j}$, an estimator of the maximal flux variation within the $j$-th time window.
We then converted each measurement $F(t)$ into a modified flux $F^*(t)$,  using the expression
$F^*(t) = \overline{F_{\rm lc}} + (F(t) - \overline{F_{\rm s}(t)}) * \sigma^2_{\rm lc} / \overline{D_j}$.
The light curve corresponding is shown in the lower panel of Fig. \ref{complete.fig}. This correction
indicates clearly that the data contain effectively a periodic component.
The period analysis was then completed using the Analysis of Variance (AoV) method of \citet{1989MNRAS.241..153S} and the TATRY code
based on the multi-harmonic periodogram of \cite{1996ApJ...460L.107S}. The value measured was $P$ = 5.5675 days. Finally, a Fourier
fit was completed according to $B_{\rm E}(t) = R_0 + \sum_{n = 1}^{n = 5} R_i \cos [(2 \pi/ P) n t + \phi_i]$. Table \ref{Fourier.tab}
reports the corresponding quantities $R_{21} = R_2 / R_1$ and $\phi_{21} = \phi_2 - 2 \phi_1$. The first row of the table refers
to the red line of the upper panel of Fig. \ref{complete.fig}. The other rows are discussed below.

\begin{figure}
	\resizebox{\hsize}{!}{\includegraphics{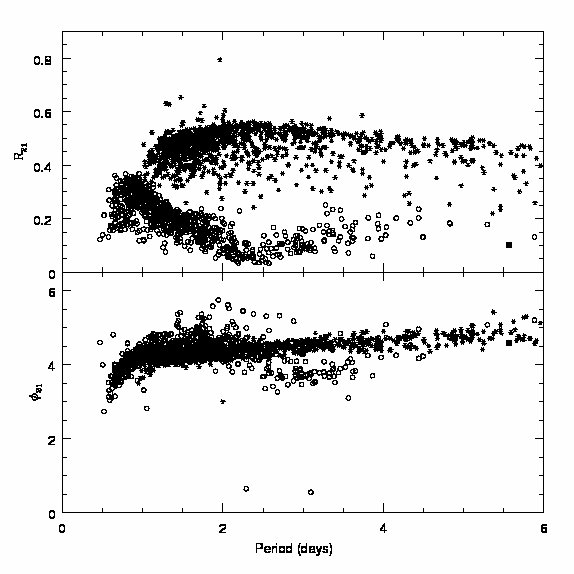}}
	\caption{$R_{21}$ and $\phi_{21}$ versus period diagrams. Stars represent the OGLE fundamental pulsators;
open circles are OGLE first overtone pulsators. Filled squares are the data on sm0060n13842 corresponding to the
first row of Table \ref{Fourier.tab}.}
	\label{FourOGLE.fig}
\end{figure}

\begin{table*}
\caption{Coefficients of the Fourier fit to the $B_{\rm E}$ light curve with a period $P$ = 5.5675 days between the epochs $E_b$ and $E_e$. $\Delta X$ represents the error in $X$. The error in $R_0$ is similar to the mean error in $B_{\rm E}$, $e.g.$ 0.029 mag, while the mean error in $R_1$ is 0.003. All other quantities are defined in the text. The first row refers to the red line in the upper panel of Fig. \ref{complete.fig}. The other rows refer to a study completed for blocks of 100 measurements, for which the long-term variation was corrected.}
\label{Fourier.tab}
\medskip
\centering

\begin{tabular}{ccccccccc}
\hline
\hline
$E_b$ - 2\,450\,000 & $E_e$ - 2\,450\,000 & $R_0$ & $R_1$ & $\phi_1$ & $R_{21}$ & $\phi_{21}$ & $\Delta R_{21}$ & $\Delta \phi_{21}$ \\
\hline
292.8648  & 2698.5268 & 16.125 & 0.082 &  0.274 & 0.101 & 4.573 & 0.022 & 0.220 \\
\hline
292.8648  & 807.6200  & 16.118 & 0.146 &  0.284 & 0.245 & 5.636 & 0.022 & 0.092 \\
488.5442  & 1037.7914 & 16.118 & 0.130 &  0.106 & 0.220 & 6.083 & 0.027 & 0.134 \\
810.6014  & 1171.5550 & 16.121 & 0.118 & -0.086 & 0.132 & 0.450 & 0.027 & 0.219 \\
1039.7866 & 1340.8647 & 16.119 & 0.095 & -0.160 & 0.026 & 1.773 & 0.030 & 1.161 \\
1173.5735 & 1430.7374 & 16.118 & 0.093 &  0.029 & 0.083 & 3.687 & 0.044 & 0.524 \\
1341.8360 & 1551.5909 & 16.122 & 0.090 &  0.398 & 0.128 & 3.542 & 0.054 & 0.414 \\
1433.7395 & 1742.8458 & 16.120 & 0.098 &  0.733 & 0.096 & 2.364 & 0.037 & 0.401 \\
1553.5609 & 1845.5713 & 16.122 & 0.099 &  0.841 & 0.134 & 2.064 & 0.036 & 0.274  \\
1744.8375 & 1924.6267 & 16.137 & 0.074 &  0.862 & 0.186 & 2.067 & 0.055 & 0.309 \\
1848.5785 & 2020.9232 & 16.132 & 0.048 &  0.763 & 0.248 & 2.516 & 0.077 & 0.335 \\
1926.6210 & 2093.7241 & 16.120 & 0.043 &  0.525 & 0.227 & 3.272 & 0.071 & 0.334 \\
2025.8887 & 2167.8788 & 16.122 & 0.047 &  0.245 & 0.217 & 4.169 & 0.066 & 0.344 \\
2093.7841 & 2223.7505 & 16.113 & 0.043 &  0.157 & 0.225 & 4.459 & 0.062 & 0.319 \\
2168.5507 & 2294.5823 & 16.113 & 0.044 &  0.557 & 0.195 & 3.862 & 0.058 & 0.299 \\
2224.7489 & 2446.7581 & 16.121 & 0.050 &  0.506 & 0.161 & 4.723 & 0.067 & 0.447 \\
2297.5690 & 2546.5336 & 16.125 & 0.057 & -0.012 & 0.158 & 0.336 & 0.056 & 0.368 \\
2448.7357 & 2622.5676 & 16.120 & 0.050 & -0.250 & 0.083 & 1.267 & 0.061 & 0.748 \\
\hline
\end{tabular}
\end{table*}
Figure \ref{FourOGLE.fig} illustrates how the fit is situated in the graphs of both $R_{21}$ and $\phi_{21}$ versus period
where the OGLE SMC Cepheids are displayed. From these graphs it appears that sm0060n13842 behaves in a similar way to a first overtone
pulsator located in the uppermost tail of period distribution of these objects (which is supported by the shape of the fit in Fig. \ref{complete.fig}). We note that the Fourier coefficients of the OGLE pulsators were measured using $I$ band data, while those for sm0060n13842 were deduced from the $B_{\rm E}$ data. We checked however, using a sample of about 1500 Cepheids cross-identified in the SMC between OGLE and EROS-2 databases, that there is a correlation with a slope of unity and a low dispersion between the OGLE and EROS-2 Fourier coefficients. From Table \ref{Photom.tab}, the Wesenheit index $W_I$ = $I$ - 1.55 $(V - I)$ \citep{1999AcA....49..437U} is equal to 13.33.  This places sm0060n13842 at the edge of the box defining first overtone Cepheids in Fig. 2 of the paper (representing the PL relation) by \citet{1999AcA....49..437U}, though in a region where few fundamental-mode objects lie.

\begin{figure}
	\resizebox{\hsize}{!}{\includegraphics{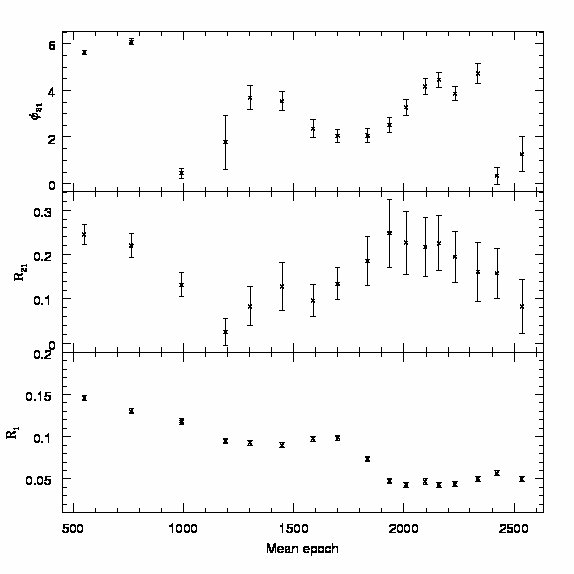}}
	\caption{Temporal evolution in the Fourier coefficients of Table \ref{Fourier.tab}. The mean epoch is the average of $E_{b}$ and $E_{e}$.}
	\label{FigFour.fig}
\end{figure}

It is evident from Fig. \ref{complete.fig} that there is significant scatter in the folded light curve.
At least in part, this is because the analysis described above is an approximation that is required to process the
data in a classical way. We examined the possibility that Fourier coefficients varied with time. We attempted to construct a light curve for which the long-term variation was corrected, but not the variable amplitude. We used a sliding window by defining blocks of 100 points every 50 measurements, with an epoch
beginning at $E_b$ and an epoch ending at $E_e$, as listed in Table \ref{Fourier.tab}. This table indicates the Fourier coefficients that
correspond to fits obtained for each block of data with the same period of 5.5675 days, which quantifies the pulsation evolution of sm0060n13842.
This analysis indicates that the $R_1$ coefficient decreases continuously with time down to a plateau, while both $R_{21}$ and $\phi_{21}$ vary with  large amplitude, as shown in Fig. \ref{FigFour.fig}. We note that $\phi_1$ also shows significant variation, which contributes to the scatter observed in the top panel of Fig. \ref{complete.fig}.

\section{Spectroscopic analysis}
\label{sec:spectro}
Low-resolution spectroscopic data were acquired using the EFOSC instrument on the ESO 3.60 m telescope. We first estimated the radial velocity (RV) by using the cross-correlation of the spectrum with a synthetic  spectrum template. We obtained  a RV of 104 km s$^{-1}$. \citet{2006AJ....131.2514H} conducted a spectroscopic survey of the SMC and measured a smooth RV distribution which was fit well by a plane. Using these data, we deduced a central RV of 145.3 km s$^{-1}$ at RA = 00:54:29 and DEC = -72:55:23, with a gradient of 5.0$\pm$0.7 km s$^{-1}$ deg$^{-1}$ East and -10.5$\pm$1.4 km s$^{-1}$ deg$^{-1}$ North. The dispersion was 26 km s$^{-1}$. Our object is located -0.22\degr ~East, 1.17\degr ~North from this central point. Hence, the predicted SMC RV, at the position of sm0010n13842, is 132.9 km s$^{-1}$ with a 1 $\sigma$ dispersion of 26 km s$^{-1}$. Thus, the observed RV is consistent, within a marging of error of 1 $\sigma$,  with this predicted value. This is supported by Fig. 9 of \citet{2006AJ....131.2514H}, which indicates that our observed RV is typical of SMC red giants.

\begin{figure}
	\resizebox{\hsize}{!}{\includegraphics{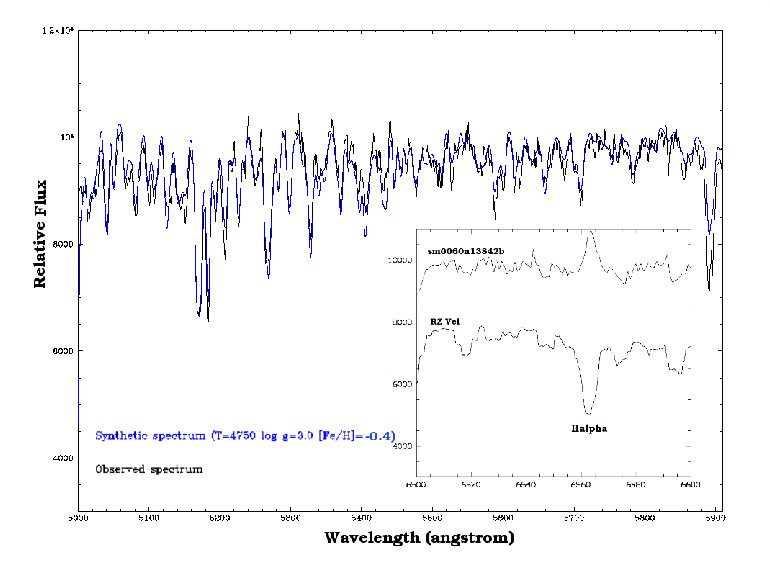}}
	\caption{EFOSC spectrum of sm0060n13842 (dark line), compared to the ``best'' synthetic spectrum (blue line). The wavelength range is 5000 to 5900 \AA{}. The insert shows the H$\alpha$ emission of sm0060n13842, compared to the absorption feature of \object{RZ Vel}.}
	\label{spectrum}
\end{figure}

As a second step, we estimated the metallicity. The critical parameter to obtain an accurate determination of metallicity is the temperature. The $(V-I)$ colour is a good indicator of the temperature of a star.  From Table \ref{Photom.tab}, we have $(V-I)$ =1.15, which gives a temperature of $\simeq 4750K$ using the synthetic colours derived from the Kurucz models \citep{1993KurCD..13.....K}. Inspection of the spectrum shows that the hydrogen line H$\alpha$ contains emission (see Fig.~\ref{spectrum}). We are therefore unable to use this line to check the temperature measured from the photometric data.

We assumed a surface gravity log~g = 3.00, which is typical of Cepheids at this period. The metallicity is relatively insensitive to this parameter. A decrease of
+1 dex in the gravity would decrease the metallicity by $\simeq$ 0.15
dex. We estimated the metallicity by comparing the observed spectrum with
a computed, synthetic spectrum (using SYNTHE) until the most appropriate match was obtained.
We also used the strong lines of the Mg triplet (located in the
5100-5200 \AA{} region) to estimate the overabundance of $\alpha$ elements.
The models were computed using the \citet{1993KurCD..13.....K} grid of models
(with the option for overshooting on). Figure~\ref{spectrum} shows the superposition of the observed spectrum
with the ``best-fit'' spectrum.  We finally obtained
a metallicity  [Fe/H]= -0.4$\pm$0.2 dex  (with [$\alpha$/Fe]$\simeq$
+0.3). This value is higher than  the mean metallicity of the SMC, -0.68 dex \citep{1998AJ....115..605L}.
It would be important to confirm this estimate with the analysis of a higher S/N and higher resolution spectrum.
As mentioned above, it is interesting that the H$\alpha$ line is detected in strong emission, while the H$\beta$ line is missing. This emission was observed in beat Cepheids by \citet{1978ApJ...226L.141B}, who noted that it could be correlated with the maximum light. To test this hypothesis, we searched for a second period in the light curve, shown on Fig. \ref{complete.fig}, by subtracting the Fourier model (given in the first row of Table \ref{Fourier.tab}) from the light curve, and reapplying the AoV method. We detected only the first (and no second) period to an accuracy of $10^{-3}$ day. Our object is therefore not a beat Cepheid. \citet{2004AJ....128.2988S} studied the H$\alpha$ emission of Galactic Cepheids and concluded that this emission can help discriminate between type I and II Cepheids in the period range 11-34 days. In the intermediate range from 3 to 8 days, they also observed that no H$\alpha$ emission was found but weak emission can occur for type II Cepheids with periods below 3 days. Additional spectroscopic data for sm0060n13842 of the entire pulsation cycle would be required to confirm these findings.

\section{The question of blending}
\label{sec:blending}
\begin{figure}
	\resizebox{\hsize}{!}{\includegraphics{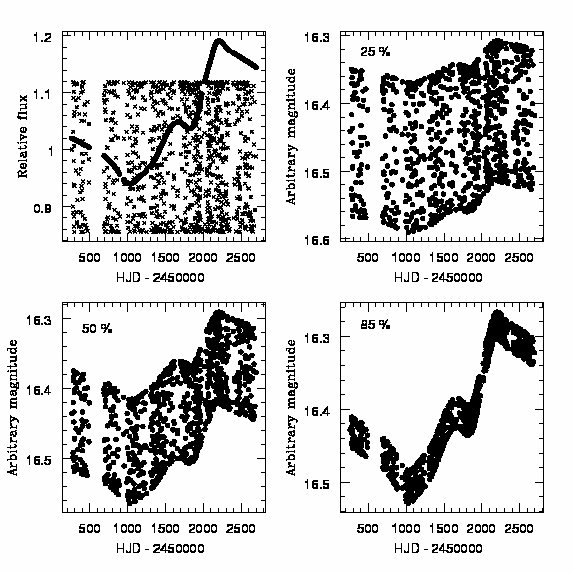}}
	\caption{Blend effect of a long-term variable object on a classical Cepheid modelled by the parameters of the second row of Table \ref{Fourier.tab} and $P$ = 5.5675 days. Top left panel: normalised fluxes of a synthetic, variable object (solid line obtained by smoothing the $B_{\rm E}$ observational data from a spline interpolation of the mean flux at mean epochs of the rows of Table \ref{Fourier.tab}) and the test Cepheid (crosses that represent the normalised flux of the test Cepheid, at the same epochs for which EROS-2 measured the flux of sm0060n13842). Other panels: light curves obtained from the sum of these two fluxes expressed on an arbitrary, magnitude scale for different relative contributions of the blend to the total flux given by the percentage on the panels.}
	\label{Blend.fig}
\end{figure}

When about a hundred million of stars are monitored in crowded fields, as in the case of the EROS-2 collaboration experiment, it is unsurprising to observe some light curves with unusual behaviour due to the blending of multiple variable stars. We ask how a (long-term) variable star blended with a (classical) Cepheid could produce the peculiar light curve presented in Fig. \ref{LightCurves.fig}. There are several arguments against this, the most important being that ascribingiveg the long-term variations to one star and the short-term variations to a constant-amplitude periodic star cannot produce the observed light curve. Such a blend would produce a flux as a function of time of the form $F(t) = F_1(t) + F_2 + A_2~\mathrm{sin} (\omega t)$, where $F_1(t)$ is the slowly varying flux of the first star and $F_2$ and $A_2$ are the time-independent mean flux and amplitude of the periodic star of frequency $\omega$. The constant $A_2$ contrasts with the observed time-dependent amplitude, which varies within an amplitude variation from $\pm$ 15\% of a magnitude 16 star to an amplitude variation of $\pm$ 5\% of a magnitude 16 star. The light curve of sm0060n13842 (Fig. \ref{LightCurves.fig}) is considerably different from those of the simulated blends shown in Fig. \ref{Blend.fig}, which provides the light curves of three reasonable choices of flux ratios for a hypothetical long-term variable and a Cepheid. The behaviour of sm0060n13842 cannot be reproduced because the test Cepheid has a constant amplitude variation in the small range of magnitude considered.

A second argument against the blending origin is that \citet{2002A&A...391..795A} showed, by studying the effects of blending on the light curve shape of Cepheids (first overtone or fundamental mode), that the maximum variation in the Fourier parameter $R_{21}$ and $\phi_{21}$ were 0.02 and 0.4, respectively, for a blend that was twice as bright as the average luminosity of a Cepheid. These values are relatively small compared to our observed variation, which is ten times larger.

\begin{figure}
	\resizebox{\hsize}{!}{\includegraphics{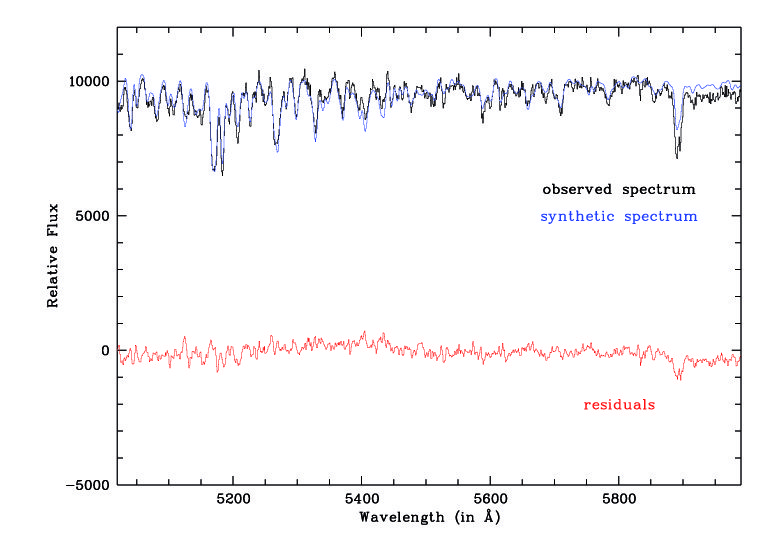}}
	\caption{Residuals from the subtraction of the synthetic Cepheid spectrum to the observed one (see Fig. \ref{spectrum}).}
	\label{residual}
\end{figure}

A final argument against blending is that the spectrum of Fig. \ref{spectrum} provides no indication of significant blending. We subtracted the Cepheid synthetic spectrum presented in Fig. \ref{spectrum} from the observed one. The result is shown in Fig. \ref{residual}. The tiny feature close to 5900 \AA{} originates in the Na D doublet. Since we assumed that [Na/Fe] = 0 in our computations, this small signal in the residuals could be explained by a non-solar Na/Fe ratio, or by a solar Na/Fe and  contamination of the spectrum by interstellar Na absorption lines. The residuals presented in Fig. \ref{residual} are similar to noise at least 20 times dimmer than the signal. This implies that a hypothetic blended companion could be fainter by the same amount.

We note that one faint star was reported by \citet{2002AJ....123..855Z} in the immediate vicinity (2.1\arcsec) of sm0060n13842, which was not found by the EROS starfinder. The star's $BVI$ magnitudes are 19.93, 19.15 and 18.11, respectively, $\sim$2.9 magnitudes fainter than sm0060n13842. This faint star could contribute a maximum of $\sim$ 7\% to the total flux, but studies completed about the behaviour of our photometric algorithm, in the case of blending, show that this value decreases linearly with the neighbour star distance. We estimate that the faint-star contribution to the sm0060n13842 flux should be approximately 2\%, which corresponds to 0.02 mag. This is comparable to the EROS-2 photometric resolution at magnitude 16. If this star was variable, we would have detected a luminosity variation at its position on the subtraction images, produced by the differential photometry (see next section). We observed no such variation apart from the luminosity variation of sm0060n13842. We note that the faint neighbour had no influence on the EFOSC spectrum, since the data were taken in 1\arcsec ~seeing.

In summary, the light curve of sm0060n13842 cannot be generated by blending a long-term variable with a constant-amplitude Cepheid. The curve can be reproduced only by a blend of a long-term variable with a variable-amplitude periodic star. Such an explanation apperas unlikely, since we would expect many more blends of long-term variable stars with constant-amplitude Cepheids and none has been found in the EROS-2 catalogue. We estimated that the probability of such an alignment of stars in the complete EROS-2 fields is less than 1\%.

\section{The question of proper motion}
\label{sec:prop_mot}
The NOMAD catalogue reports a proper motion measurement for sm0060n13842
 of $12\pm7$ mas.yr$^{-1}$ along right ascension and $10\pm4$ mas.yr$^{-1}$ along
 declination. These small positive values are consistent with no proper motion
 at the 1.5 and 2.5 $\sigma$ significance level for right ascension and declination, respectively. We used the EROS-2
 images to measure these values where detectable.

\begin{figure}
	\resizebox{\hsize}{!}{\includegraphics{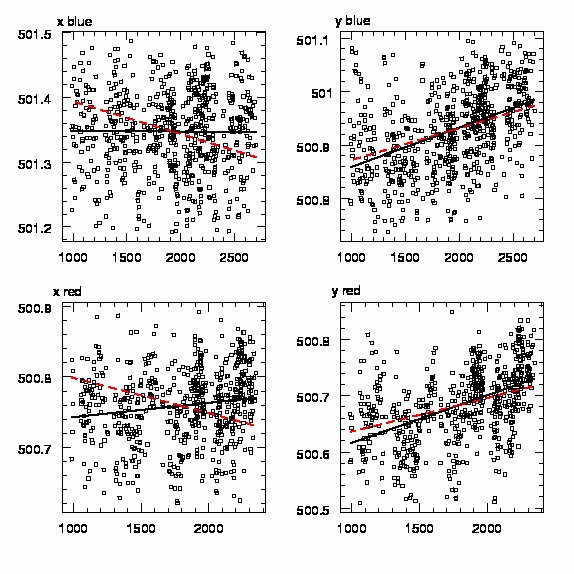}}
	\caption{Temporal evolution of the pixel coordinates of sm0060n13842. Abscissas are HJD - 2\,450\,000. Regression lines are indicated in solid black, while dashed, red ones show proper motion given by the NOMAD catalogue. The $x$ coordinate increases with decreasing declination ($e.g.$ towards South), while the $y$ coordinate increases with increasing right ascension.}
	\label{PropMot}
\end{figure}

The ISIS image subtraction package (Alard 2000) was used to geometrically align
 a set of 1K x 1K sub-images centered on sm0060n13842. The template images,
 used to measure both colours, were compiled from 20 images corresponding to the highest quality seeing conditions. After 
substraction, the centroid coordinates of the peculiar variable object were
 measured in each image. These coordinate distributions were clipped at
 2 $\sigma$ significance level to be conservative and to reduce the sensitivity to
 systematic errors. Figure \ref{PropMot} shows, in both colours and both directions, the
 temporal evolution of the pixel coordinates of sm0060n13842 and the
 linear-regression lines (black solid lines) compared to the proper motion
 reported by the NOMAD catalogue (red dashed lines). Table \ref{propmot.tab} summarizes the
 measurements.
 
The slopes measured on both mosaics are in excellent agreement with each other.
 Our main signal is along the $y$ axis (\textit{i.e.} right ascension). We did not extensively study all possible systematic effects on our astrometric data, but another
 method was used to verify this proper motion. It is based on a 1D
 PSF fit along both axes completed during the EROS-2 photometric process, when images
 were geometrically aligned to a reference image. The new proper-motion measurements
 for our object agree with the values listed in Table \ref{propmot.tab}; an interesting
 result is produced the thousand bright stars surrounding sm0060n13842.
 With them, we estimated that a proper motion of about 17 mas.yr$^{-1}$, along
 the right ascension, was significant only at 3 $\sigma$ level.
 Therefore we point out that the uncertainty in our measurement is underestimated, and that
 a systematic error of up to $\sim6$ mas.yr$^{-1}$ should be added. The measured, 
 mean proper motion is therefore $17.4\pm6.0$ mas.yr$^{-1}$ along right ascension, and 
 $-2.4\pm6.0$ mas.yr$^{-1}$ along declination. These values are consistent with the 
 measurement reported in NOMAD, but do not confirm the highest NOMAD signal at a
 2.5 $\sigma$ significance level along the declination axis.

If the positive proper motion signal found along right ascension is confirmed,
 sm0060n13842 is a candidate for being a foreground Galactic star. Furthermore,
 we note that a proper motion towards the increasing direction of right ascension is as expected for a
 Galactic star in the direction of the SMC. More disturbing is the result
 obtained from the reduced proper motion (RPM). In Fig. 2 of \citet{2002ApJ...575L..83S},
 sm0060n13842 is located (with $V-J\sim2$ and RPM $\sim7.4$)\footnote{These values were calculated 
 using magnitudes listed in Table \ref{Photom.tab}.
 RPM = $V + 5$ log$\mu$, where $\mu$ is the proper motion expressed in "yr$^{-1}$}
 in the dense main-sequence star region. Therefore, the measured proper motion
 (value and direction) of our peculiar SMC Cepheid-like object is surprisingly
 consistent with those measured for nearby main-sequence stars.

\begin{table}
	\caption{Linear regression coefficients corresponding to the solid lines in Fig. \ref{PropMot}.}
	\label{propmot.tab}
	\medskip
	\centering
	\begin{tabular}{ccc}
	\hline
	\hline
	Coordinate & Slope                                            & Proper motion \\
	                    & (10$^{-5}$ pixel day$^{-1}$) &  (mas yr$^{-1}$) \\
	\hline
	$x$ blue		& 0.12 $\pm$ 0.53 &  -0.3  $\pm$ 1.2 \\
	$x$ red		& 1.92 $\pm$ 0.55 &  -4.2 $\pm$ 1.2 \\
	$y$ blue		& 7.59 $\pm$ 0.58 & 16.6 $\pm$ 1.3 \\
	$y$ red		& 8.33 $\pm$ 0.63 & 18.2 $\pm$ 1.4 \\
	\hline
	\end{tabular}
\end{table}

\section{The post-AGB hypothesis}
\label{sec:post-AGB}
Since it is known that the post-AGB tracks cross the high-luminosity end of the Population II Cepheids instability strip \citep{2003ARA&A..41..391V}, we examine the facts that argue against  sm0060n13842 being a post-AGB object. First, such stars often exhibit a near-IR excess. In the LMC, $J - K$  is in the range 2 - 4 for MSX sources \citep{2001pao..conf...71W} and in the range  6 - 8 for the Arecibo sample of OH/IR stars \citep{2001pao..conf...49J}. For the present case the mean, $J - K$ colour index is 0.825 from Table \ref{Photom.tab}. In a similar way, no source was found in the IRAS catalogue in the vicinity of our object.  However, these arguments should be interpreted with caution because of the low value of gas-dust ratio in the SMC. Second, it is recognized that the class of RV Tauri stars harbors post-AGB objects, although the shortest periods in this class are in the range of a few tens of days \citep{2002PASP..114..689W}, far longer than the period of  sm0060n13842. In addition, a pulsation model has been used (Wood, private communication) with the following parameters derived from the $V_0$ and $(V - I)_0$ data: $L = 1660 L_{\sun}$; $T_{eff}$ = 4500 K. With that luminosity, a post-AGB star would have a mass of $\sim$0.52 $M_{\sun}$, and a core mass very slightly smaller ($\sim$0.517 $M_{\sun}$ for the derived $T_{eff}$). Computing the periods for such a star,  we find that P$_{0} \sim$ 65 days and P$_1 \sim$ 20 days, which implies that both models ARE very unstable. This rules out the interpretation of a SMC post-AGB object.

\section{Conclusion}
The nature of sm0060n13842 remains uncertain, according to the different points discussed above:
\begin{itemize}
 \item The period of 5.5675 days, the apparent magnitude, and the Fourier coefficients indicate that sm0060n13842 behaves like a Cepheid located in the SMC. However, we have no clear indication from Figs. \ref{CMD.fig} and \ref{FourOGLE.fig} that this star belongs to the class of either fundamental or first overtone pulsators. The Fourier coefficients show a temporal evolution that, to our knowledge, was not previously observed (we have excluded that these properties can be explained by the blending of a Cepheid and a longer timescale variable star). 
\item Our simple spectroscopic analyses measures a radial velocity that is consistent with spectroscopic survey of the SMC completed by \citet{2006AJ....131.2514H}. However, the derived metallicity is higher than typical values measured for the SMC.
\item The hypothesis of a blend is unlikely because the peculiar evolution in the amplitude with time cannot be explainded by a blend with a long-term variable. Moreover, the subtraction of a synthetic, Cepheid spectrum from the observed spectrum does not provide any significant residuals.
\item Considering the position of sm0060n13842 on EROS images, the possibility of a proper motion in right ascension cannot be ruled out. This is partially consistent with measurements reported in the NOMAD catalogue. A proper motion of 17 mas yr$^{-1}$, at a distance of 60 kpc, produces a very high transverse velocity of $\sim$4800 km s$^{-1}$, suggesting that the object was ejected from the SMC. If \textit{a contrario} we assume that sm0060n13842 is a Galactic object, it is located on the main sequence. In such a case however, we would have to explain why its behavior is that of a Cepheid at the distance of the SMC.
\end{itemize}
This discussion indicates that sm0060n13842 is an extremely interesting object which should be reobserved more thoroughly, both photometrically and
spectroscopically. In this respect, the OGLE-III project data should certainly provide invaluable information.

\begin{acknowledgements}
We are grateful to the PLANET collaboration for access to the 1m54 Danish telescope at ESO during the August 2004 observation campaign. JPB, JBM and ASC acknowledge additional travel support from the LEA Astro-PF fund. JBM and PT thank V. Hill for valuable suggestions and the Collaboration thanks P. Wood for his post-AGB computations. This publication makes use of data products from the Two Micron All Sky Survey, which is a joint project of the University of Massachusetts and the Infrared Processing and Analysis Center/California Institute of Technology, funded by the National Aeronautics and Space Administration and the National Science Foundation. The DENIS project has been partly funded by the SCIENCE and the HCM plans of the European Commission under grants CT920791 and CT940627. It is supported by INSU, MEN and CNRS in France, by the State of Baden-Württemberg in Germany, by DGICYT in Spain, by CNR in Italy, by FFwFBWF in Austria, by FAPESP in Brazil, by OTKA grants F-4239 and F-013990 in Hungary, and by the ESO CSP in Brazil, and by the ESO C\&EE grant A-04-046. This research has made use of the VizieR catalogue access tool, CDS, Strasbourg, France. This research has made use of the NASA/IPAC Extragalactic Database (NED) which is operated by the Jet Propulsion Laboratory, California Institute of Technology, under contract with the National Aeronautics and Space Administration.
\end{acknowledgements}

\bibliographystyle{aa}
\bibliography{6915main}

\begin{thebibliography}{40}
\expandafter\ifx\csname natexlab\endcsname\relax\def\natexlab#1{#1}\fi

\bibitem[{{Ansari}(1996)}]{1996VA.....40..519A}
{Ansari}, R. 1996, Vistas in Astronomy, 40, 519

\bibitem[{{Antonello}(2002)}]{2002A&A...391..795A}
{Antonello}, E. 2002, \aap, 391, 795

\bibitem[{{Barrell}(1978)}]{1978ApJ...226L.141B}
{Barrell}, S.~L. 1978, \apjl, 226, L141

\bibitem[{{Burki} \& {Mayor}(1980)}]{1980A&A....91..115B}
{Burki}, G. \& {Mayor}, M. 1980, \aap, 91, 115

\bibitem[{{Burki} {et~al.}(1982){Burki}, {Mayor}, \&
  {Benz}}]{1982A&A...109..258B}
{Burki}, G., {Mayor}, M., \& {Benz}, W. 1982, \aap, 109, 258

\bibitem[{{Cioni} {et~al.}(2000){Cioni}, {Loup}, {Habing}, {Fouqu{\'e}},
  {Bertin}, {Deul}, {Egret}, {Alard}, {de Batz}, {Borsenberger}, {Dennefeld},
  {Epchtein}, {Forveille}, {Garz{\'o}n}, {Hron}, {Kimeswenger}, {Lacombe}, {Le
  Bertre}, {Mamon}, {Omont}, {Paturel}, {Persi}, {Robin}, {Rouan}, {Simon},
  {Tiph{\`e}ne}, {Vauglin}, \& {Wagner}}]{2000A&AS..144..235C}
{Cioni}, M.-R., {Loup}, C., {Habing}, H.~J., {et~al.} 2000, \aaps, 144, 235

\bibitem[{{Cutri} {et~al.}(2003){Cutri}, {Skrutskie}, {van Dyk}, {Beichman},
  {Carpenter}, {Chester}, {Cambresy}, {Evans}, {Fowler}, {Gizis}, {Howard},
  {Huchra}, {Jarrett}, {Kopan}, {Kirkpatrick}, {Light}, {Marsh}, {McCallon},
  {Schneider}, {Stiening}, {Sykes}, {Weinberg}, {Wheaton}, {Wheelock}, \&
  {Zacarias}}]{2003yCat.2246....0C}
{Cutri}, R.~M., {Skrutskie}, M.~F., {van Dyk}, S., {et~al.} 2003, VizieR Online
  Data Catalog, 2246, 0

\bibitem[{{Deasy} \& {Wayman}(1985)}]{1985MNRAS.212..395D}
{Deasy}, H.~P. \& {Wayman}, P.~A. 1985, \mnras, 212, 395

\bibitem[{{Denis Consortium}(2005)}]{2005yCat....102002T}
{Denis Consortium}, T. 2005, VizieR Online Data Catalog, 1, 2002

\bibitem[{{Derue} {et~al.}(2001){Derue}, {Afonso}, {Alard}, {Albert},
  {Andersen}, {Ansari}, {Aubourg}, {Bareyre}, {Bauer}, {Beaulieu}, {Blanc},
  {Bouquet}, {Char}, {Charlot}, {Couchot}, {Coutures}, {Ferlet}, {Fouqu{\'e}},
  {Glicenstein}, {Goldman}, {Gould}, {Graff}, {Gros}, {Ha{\"\i}ssinski},
  {Hamilton}, {Hardin}, {de Kat}, {Kim}, {Lasserre}, {Le Guillou}, {Lesquoy},
  {Loup}, {Magneville}, {Mansoux}, {Marquette}, {Maurice}, {Milsztajn},
  {Moniez}, {Palanque-Delabrouille}, {Perdereau}, {Pr{\'e}vot}, {Regnault},
  {Rich}, {Spiro}, {Vidal-Madjar}, {Vigroux}, \&
  {Zylberajch}}]{2001A&A...373..126D}
{Derue}, F., {Afonso}, C., {Alard}, C., {et~al.} 2001, \aap, 373, 126

\bibitem[{{Derue} {et~al.}(2002){Derue}, {Marquette}, {Lupone}, {Afonso},
  {Alard}, {Albert}, {Amadon}, {Andersen}, {Ansari}, {Aubourg}, {Bareyre},
  {Bauer}, {Beaulieu}, {Blanc}, {Bouquet}, {Char}, {Charlot}, {Couchot},
  {Coutures}, {Ferlet}, {Fouqu{\'e}}, {Glicenstein}, {Goldman}, {Gould},
  {Graff}, {Gros}, {Ha\&{\i}ssinski}, {Hamilton}, {Hardin}, {de Kat}, {Kim},
  {Lasserre}, {Le Guillou}, {Lesquoy}, {Loup}, {Magneville}, {Mansoux},
  {Maurice}, {Milsztajn}, {Moniez}, {Palanque-Delabrouille}, {Perdereau},
  {Pr{\'e}vot}, {Regnault}, {Rich}, {Spiro}, {Vidal-Madjar}, {Vigroux},
  {Zylberajch}, \& {The EROS collaboration}}]{2002A&A...389..149D}
{Derue}, F., {Marquette}, J.-B., {Lupone}, S., {et~al.} 2002, \aap, 389, 149

\bibitem[{{Eddington} \& {Plakidis}(1929)}]{1929MNRAS..90...65E}
{Eddington}, A.~S. \& {Plakidis}, S. 1929, \mnras, 90, 65

\bibitem[{{Gilliland} \& {Fisher}(1985)}]{1985PASP...97..285G}
{Gilliland}, R.~L. \& {Fisher}, R. 1985, \pasp, 97, 285

\bibitem[{{Hamadache} {et~al.}(2006){Hamadache}, {Le Guillou}, {Tisserand},
  {Afonso}, {Albert}, {Andersen}, {Ansari}, {Aubourg}, {Bareyre}, {Beaulieu},
  {Charlot}, {Coutures}, {Ferlet}, {Fouqu{\'e}}, {Glicenstein}, {Goldman},
  {Gould}, {Graff}, {Gros}, {Haissinski}, {de Kat}, {Lesquoy}, {Loup},
  {Magneville}, {Marquette}, {Maurice}, {Maury}, {Milsztajn}, {Moniez},
  {Palanque-Delabrouille}, {Perdereau}, {Rahal}, {Rich}, {Spiro},
  {Vidal-Madjar}, {Vigroux}, \& {Zylberajch}}]{2006A&A...454..185H}
{Hamadache}, C., {Le Guillou}, L., {Tisserand}, P., {et~al.} 2006, \aap, 454,
  185

\bibitem[{{Harris} \& {Zaritsky}(2006)}]{2006AJ....131.2514H}
{Harris}, J. \& {Zaritsky}, D. 2006, \aj, 131, 2514

\bibitem[{{Hubble}(1926)}]{1926ApJ....63..236H}
{Hubble}, E.~P. 1926, \apj, 63, 236

\bibitem[{{Jim{\'e}nez-Esteban} {et~al.}(2001){Jim{\'e}nez-Esteban}, {Engels},
  \& {Garc{\'{\i}}a-Lario}}]{2001pao..conf...49J}
{Jim{\'e}nez-Esteban}, F., {Engels}, D., \& {Garc{\'{\i}}a-Lario}, P. 2001, in
  Post-AGB Objects as a Phase of Stellar Evolution, ed. R.~{Szczerba} \& S.~K.
  {G{\'o}rny}, p. 49

\bibitem[{{Koen}(2001)}]{2001MNRAS.322...97K}
{Koen}, C. 2001, \mnras, 322, 97

\bibitem[{{Kurucz}(1993)}]{1993KurCD..13.....K}
{Kurucz}, R. 1993, ATLAS9 Stellar Atmosphere Programs and 2 km/s grid.~Kurucz
  CD-ROM No.~13.~ Cambridge, Mass.: Smithsonian Astrophysical Observatory,
  1993., 13

\bibitem[{{Luck} {et~al.}(1998){Luck}, {Moffett}, {Barnes}, \&
  {Gieren}}]{1998AJ....115..605L}
{Luck}, R.~E., {Moffett}, T.~J., {Barnes}, III, T.~G., \& {Gieren}, W.~P. 1998,
  \aj, 115, 605

\bibitem[{{Macri} {et~al.}(2001){Macri}, {Sasselov}, \&
  {Stanek}}]{2001ApJ...550L.159M}
{Macri}, L.~M., {Sasselov}, D.~D., \& {Stanek}, K.~Z. 2001, \apjl, 550, L159

\bibitem[{{Paczynski}(1986)}]{1986ApJ...304....1P}
{Paczynski}, B. 1986, \apj, 304, 1

\bibitem[{{Percy} \& {Colivas}(1999)}]{1999PASP..111...94P}
{Percy}, J.~R. \& {Colivas}, T. 1999, \pasp, 111, 94

\bibitem[{{Percy} {et~al.}(2003){Percy}, {Velocci}, \&
  {Sterken}}]{2003PASP..115..626P}
{Percy}, J.~R., {Velocci}, V., \& {Sterken}, C. 2003, \pasp, 115, 626

\bibitem[{{Pietrukowicz}(2001)}]{2001AcA....51..247P}
{Pietrukowicz}, P. 2001, Acta Astronomica, 51, 247

\bibitem[{{Pietrukowicz}(2002)}]{2002AcA....52..177P}
{Pietrukowicz}, P. 2002, Acta Astronomica, 52, 177

\bibitem[{{Pietrukowicz}(2003)}]{2003AcA....53...63P}
{Pietrukowicz}, P. 2003, Acta Astronomica, 53, 63

\bibitem[{{Saitou} \& {Takeuti}(1990)}]{1990PASJ...42..341S}
{Saitou}, M. \& {Takeuti}, M. 1990, \pasj, 42, 341

\bibitem[{{Salim} \& {Gould}(2002)}]{2002ApJ...575L..83S}
{Salim}, S. \& {Gould}, A. 2002, \apjl, 575, L83

\bibitem[{{Schmidt} {et~al.}(2004){Schmidt}, {Johnston}, {Lee}, {Langan},
  {Newman}, \& {Snedden}}]{2004AJ....128.2988S}
{Schmidt}, E.~G., {Johnston}, D., {Lee}, K.~M., {et~al.} 2004, \aj, 128, 2988

\bibitem[{{Schwarzenberg-Czerny}(1989)}]{1989MNRAS.241..153S}
{Schwarzenberg-Czerny}, A. 1989, \mnras, 241, 153

\bibitem[{{Schwarzenberg-Czerny}(1996)}]{1996ApJ...460L.107S}
{Schwarzenberg-Czerny}, A. 1996, \apjl, 460, L107

\bibitem[{{Tisserand} {et~al.}(2007){Tisserand}, {Le Guillou}, {Afonso},
  {Albert}, {Andersen}, {Ansari}, {Aubourg}, {Bareyre}, {Beaulieu}, {Charlot},
  {Coutures}, {Ferlet}, {Fouqu{\'e}}, {Glicenstein}, {Goldman}, {Gould},
  {Graff}, {Gros}, {Haissinski}, {Hamadache}, {de Kat}, {Lasserre}, {Lesquoy},
  {Loup}, {Magneville}, {Marquette}, {Maurice}, {Maury}, {Milsztajn}, {Moniez},
  {Palanque-Delabrouille}, {Perdereau}, {Rahal}, {Rich}, {Spiro},
  {Vidal-Madjar}, {Vigroux}, {Zylberajch}, \& {The EROS-2
  Collaboration}}]{2007A&A...469..387T}
{Tisserand}, P., {Le Guillou}, L., {Afonso}, C., {et~al.} 2007, \aap, 469, 387

\bibitem[{{Turner} {et~al.}(1999){Turner}, {Horsford}, \&
  {MacMillan}}]{1999JAVSO..27....5T}
{Turner}, D.~G., {Horsford}, A.~J., \& {MacMillan}, J.~D. 1999, Journal of the
  American Association of Variable Star Observers (JAAVSO), 27, 5

\bibitem[{{Udalski} {et~al.}(1999){Udalski}, {Soszynski}, {Szymanski},
  {Kubiak}, {Pietrzynski}, {Wozniak}, \& {Zebrun}}]{1999AcA....49..437U}
{Udalski}, A., {Soszynski}, I., {Szymanski}, M., {et~al.} 1999, Acta
  Astronomica, 49, 437

\bibitem[{{Van Hoolst} \& {Waelkens}(1995)}]{1995A&A...295..361V}
{Van Hoolst}, T. \& {Waelkens}, C. 1995, \aap, 295, 361

\bibitem[{{van Winckel}(2003)}]{2003ARA&A..41..391V}
{van Winckel}, H. 2003, \araa, 41, 391

\bibitem[{{Wallerstein}(2002)}]{2002PASP..114..689W}
{Wallerstein}, G. 2002, \pasp, 114, 689

\bibitem[{{Wood} \& {Cohen}(2001)}]{2001pao..conf...71W}
{Wood}, P.~R. \& {Cohen}, M. 2001, in Post-AGB Objects as a Phase of Stellar
  Evolution, ed. R.~{Szczerba} \& S.~K. {G{\'o}rny}, p. 71

\bibitem[{{Zaritsky} {et~al.}(2002){Zaritsky}, {Harris}, {Thompson}, {Grebel},
  \& {Massey}}]{2002AJ....123..855Z}
{Zaritsky}, D., {Harris}, J., {Thompson}, I.~B., {Grebel}, E.~K., \& {Massey},
  P. 2002, \aj, 123, 855

\end{thebibliography}

\end{document}